\documentclass[letterpaper, 10 pt, conference]{ieeeconf}  
\usepackage{microtype}
\usepackage{graphicx}
\usepackage{subfigure}
\usepackage{booktabs} 
\usepackage{amsmath}
\usepackage{amssymb}
\usepackage{tabularx}
\usepackage{multirow}
\usepackage{hyperref}
\usepackage{adjustbox}
\usepackage{cite}
\usepackage{soul}
\usepackage[colorinlistoftodos]{todonotes}

\IEEEoverridecommandlockouts                              

\title{\large \bf No Transfers Required: Integrating Last Mile with Public Transit Using Opti-Mile}

\author{Raashid Altaf$^{A}$ and Pravesh Biyani$^{B}$
\thanks{$^{A}$Raashid Altaf is a PhD Candidate in Department of Computer Science and Engineering,
        Indraprashta Institute of Information Technology, Delhi, India
        {\tt\small raashida@iiitd.ac.in}}%
\thanks{$^{B}$Pravesh Biyani is an Associate Professor with the Department of Electronics and Communications Engineering, Indraprashta Institute of Information Technology, Delhi, India
        {\tt\small praveshb@iiitd.ac.in}}%
}

\begin{document}

\maketitle
\thispagestyle{empty}
\pagestyle{empty}

\begin{abstract}
\par Public transit is a popular mode of transit due to its affordability, despite the inconveniences due to the necessity of transfers required to reach most areas.  For example, in the bus and metro network of New Delhi, only 30\% of stops can be directly accessed from any starting point, thus requiring transfers for most commutes. Additionally, first/last mile services like rickshaws, tuk-tuks or shuttles are commonly used as feeders to the nearest public transit access points, which leads to sub-par utilisation of the transit network. Ultimately, users often face a trade-off between coverage and transfers to reach their destination, regardless of the mode of transit. 
\par To address the problem of limited accessibility and inefficiency due to transfers in public transit systems, we propose ``opti-mile," a novel trip planning approach that combines first/last mile services with public transit such that no transfers are required\footnote{Our work is based on the assumption that the transfers from first/last mile are necessary for any public transit commute, whether it be through walking or taking a first/last mile service. Thus any further mention of transfers specifically refers to transfers within public transit, unless otherwise mentioned}. Opti-mile allows users to customise trip parameters such as maximum walking distance, and acceptable fare range. We analyse the transit network of New Delhi, evaluating the efficiency, feasibility and advantages of opti-mile multi-modal trips between randomly selected source-destination pairs. We demonstrate that opti-mile trips lead to a 10\% reduction in distance travelled for 18\% increase in average price compared to traditional shortest paths using just public transit. We also show that opti-mile trips provide better coverage of the city than public transit, without a significant fare increase.
\end{abstract}
\section{Introduction}
\label{intro}
\par Public transportation is a cost-efficient mode of travel. However, in developing countries like India, it may not always be the most reliable option due to delays and other operational issues.  Rapid urbanisation in these regions has led to escalating private vehicle usage, resulting in typical transportation issues such as traffic, crowded public transit, and pollution. Such problems are further exacerbated by unplanned residential expansion.
\par Although public transit often provides broad coverage across the serviced region, achieving it often involves transfers between routes or different modes of transit. In Delhi, for example, only about 30\% of the stops can be directly accessed from any starting stop, without any transfers. Transfers in public transit are inconvenient and introduce uncertainty into travel plans due to potential delays. Moreover, the extent of coverage provided by public transit heavily relies on the willingness of users to walk significant distances. 
\par Transit coverage in a specific area refers to the percentage of the total area that you can reach within a set distance of $k$ kilometers from the nearest transit access point. For example, within New Delhi, public transit encompasses 97.40\% of the area of the most densely populated municipal wards, considering a radius of 1 kilometer. However, this coverage reduces by up to 35\% when restricted to a 500-meter radius due to limited direct connectivity among the residential and commercial centers. The mitigation of this coverage gap often necessitates the usage of first/last mile (LM) services.
\par The first/last mile refers to the initial/final leg of a person's journey from a transit stop to their destination and is often the most critical part of the commute. First/Last mile services are of two types: dedicated first/last mile, which includes bike/car taxis and shared first/last mile, which include rickshaws, \textit{tuk tuk} and other modes of transportation common in South-East Asia and other developing countries worldwide as a cheap and efficient way to get around a city.
\par While the first/last mile is an obvious way to increase the coverage provided by public transit, it is often just used as a feeder to the nearest public transit access point. Most people walk or take a first/last mile service to the nearest stop, as it makes the most intuitive sense. However, adding a first/last mile leg to the public transit adds a layer of complexity to the trip where a user often has to make a trade-off between the convenience of a first/last mile service and the number of transfers they have to take to reach their destination. There is seemingly only the option of either taking an end-to-end cab or enduring the transfers and delays associated with PT, which decreases the efficiency of the trip. \textit{In this paper, we posit that a trip that includes a first/last mile and no transfers in the public transit leg is a more efficient way to travel.}
\subsection{Opti-Mile}
\par
\begin{figure}[b]
    \centering
    \includegraphics[width = \columnwidth]{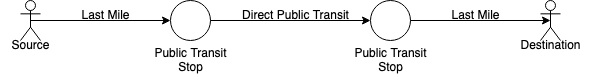}
    \caption{An Opti-Mile trip ensures a direct public transit connection with the help of first/last mile services}
    \label{optimile}
\end{figure}

 This paper introduces the term ``opti-mile" to describe a scenario where a commuter decides to walk or use a first/last mile service in order to reach a stop that provides a direct public transit (PT) connection to their destination (Fig. \ref{optimile}). By paying a slightly higher fare than a public transit trip, yet still lower than the cost of a complete taxi ride, the Opti-Mile approach enhances the overall travel experience by strategically selecting a stop that offers a more efficient public transit route, rather than simply opting for the nearest stop to the commuter's starting point.

\par We quantify efficiency of a trip as a function of convenience and cost-effectiveness of a path. Convenience measures the ease and comfort of using a transportation option, while cost-effectiveness evaluates the economic efficiency of the path in terms of value provided to the passenger. The comprehensive definitions of these  metrics can be found in Section \ref{exp1: def}
\par To implement Opti-Mile trip planning, we solve an optimisation problem that encompasses user monetary preferences along with other criteria to solve a minimisation problem that optimises the weighted travel time of a trip. (Section \ref{subsub: opt-prob}). 
\par To observe the attributes of optimal paths across various combinations of input parameters, we perform the following experiments. This analysis yields a quantification of path efficiency based on the resulting output parameters:
\begin{enumerate}
    \item We randomly sample 1000 source-destination pairs from bounding boxes selected based on the population density of the different municipal wards in the city. For each pair, we solve the optimisation problem in Section \ref{subsub: opt-prob} and observe the attributes of paths having the highest and the lowest efficiency. 
    \item We also measure the coverage provided by opti-mile trips by randomly sampling 1000 locations and measuring the area coverage achievable from the location through just opti-mile trips.
\end{enumerate}
\subsection{Contributions and Insights}
\par Main objectives in this paper are:
\begin{enumerate}
    \item To showcase opti-mile as an alternative to commute using cabs and an effective strategy for enhancing the accessibility and usability of public transit networks. We take Delhi as our case study.
    \item To highlight the effect of user commuting preferences such as fare, number of transfers, or preference of public transit over the first/last mile on the optimal path in a multi-modal transit network.
\end{enumerate}
\par Based on our experiments, we argue that the opti-mile trip is a more efficient way of integrating public transit vis-a-vis traditional LM.  Our results indicate that:
\begin{enumerate}
    \item Opti-Mile paths are 90\% more likely to be the most efficient, favoring longer first/last mile ranges over public transit transfers.
    \item An opti-mile trip of two kilometre first/last mile range provides similar area coverage as the PT-only network coverage with a 500-metre walk.
    \item Opti-Mile trips reduce travel distance by 10\% but increase fares by 18\% compared to unrestricted trips. This could attract non-regular public transit users and boost adoption of public transit.
\end{enumerate}
\section{Related Works}
\label{review}
\par Integrating first- and first/last mile services with a main transit mode has long been a crucial research topic in Operations Research. Previous studies have mainly focused on improving the efficiency and sustainability of first/last mile delivery for goods \cite{bosona_urban_2020, silva_sustainable_2023} or identifying the factors that influence the demand and satisfaction of first/last mile services for passengers 
\cite{rahman_first-and-last-mile_2022}
However, Silva \cite{silva_sustainable_2023} points out that currently there is no ideal solution for the first/last mile problem in practice, especially when considering the conflicting objectives of different stakeholders.
\par Existing approaches for solving the first/last mile problem in public transit can be classified into three categories: scheduled services \cite{grahn_optimizing_2022,xiong_optimal_2021,xie_solution_2010}, flexible and curb-to-curb services \cite{pei_flexible_2019}, or a hybrid of both \cite{qiu_demi-flexible_2015}. These approaches are variations of the Transit Route Network Design Problem (TRNDP), where the routes and frequencies of first/last mile services are optimised according to the demand at the nodes and the availability of vehicles \cite{iliopoulou_metaheuristics_2019}. Grahn \cite{grahn_optimizing_2022} proposed a system that integrates public transit shuttles with private transportation network companies like Uber to connect riders to mainline transit. Xiong \cite{xiong_optimal_2021} and Xie \cite{xie_solution_2010} design a community shuttle system that provides flexible mobility to public transit passengers to access metro/rapid transit networks. Pei \cite{pei_flexible_2019} developed a flexible routing model that offers alternative options to commuters based on their willingness to pay, aiming to balance the goals of passengers and bus companies. Authors in \cite{kumar_algorithm_2021,shaheen_mobility_2016,jain_improved_2019} aim to reduce the vehicle-hours in the system and increase transit ridership by providing a fast and reliable mobility service at a low cost.
\par Many research works on first/last mile integration concentrates on providing access to public transit rather than optimising the entire journey. This approach may improve the connectivity of shuttle services to public transit stops, but it does not guarantee that public transit is used efficiently. Our experiments show that connecting users to the nearest public transit stops often result in more transfers, leading to inefficient trips and low satisfaction with public transit \cite{park_first-first/last mile_2021}.
\par We consider the entire trip as a unit of analysis and suggest policy changes that would improve the end-to-end journey for a commuter instead of focusing on the first/last mile and public transit legs separately.
\par This work is a part of the multi-modal trip planning solution provided by the Chartr App\cite{chartr}, available on Google and iOS App Stores. Our platform can seamlessly integrate multiple last-mile/taxi service providers and suggest travel plans based on the user's chosen budget for fare, giving users a smooth and personalised door-to-door commuting experience.
\section{Trip Planning Overview}
\label{sec: TP Overview}
\par Trip Planning is a method of providing an optimal path between two points in a transit network where optimality is defined under some pre-defined contraints, e.g. travel time. Opti-mile is a trip planning method with constraints on number of transfers and fare. To explain opti-mile, we first provide a general overview of the trip planning paradigm and describe the trip planning model used.

\subsection{Trip Planning in Public Transit}
\par A trip planning system consists of two parts: a network model, and an optimisation problem. 
\subsubsection{Network Model} \label{subsub:net-model} We model the transit network as a graph G (V, E) such that V denotes a set $\{v_1, v_2, \ldots v_n\}$ of public transit stops and $E = \{e_1, e_2, \ldots, e_m\}$ denotes the weighted edges between any two stops having weight $w_i \forall e_i \in E$. We also define a set of $|R|$ routes indexed by unique IDs $R\subset N$.
\par Each route  $r \in R$ is defined as a sequence of edges $\{e_{i_1}, e_{i_2}, \dots, e_{i_k}\}$ for $e_{i_j} \in E$, $i_j \in \{1, 2, \ldots, m\}$, and $j \in \{1, 2, ..., k\}$.
\subsubsection{Cost of a Path} \label{subsub: path cost}The cost of a path in a graph is the sum of edge weights. In case of trip planning, the edge-weight is the travel time between two nodes. Let $p(v_i, v_j) = \{e_{b_1}, e_{b_2}, \dots, e_{b_a}\}$ be a sequence of $a$ edges from $v_i$ to $v_j$ with edge-weights $\{w_{b_1}, w_{b_2}, \dots, w_{b_a}\}$ respectively. The cost $c(p)$ of the path $p(v_i, v_j)$ then defined as:

\begin{equation}
    c(p) = \sum_{e_{b_k} \in p\left(v_i, v_j\right)} w_{b_k}
\end{equation}
\subsubsection{Optimisation Problem} The shortest path problem is the most commonly used method to obtain the optimal path in a transit network graph. If $P\left(u, v\right) = \left\{p_1\left(u, v\right), p_2\left(u, v\right), \ldots, p_t\left(u, v\right)\right\}$ is the set of $t$ possible paths between a source-destination pair $\left(u, v\right)$, the optimal path is a path $\tilde{p}\left(u, v\right)$ such that;
\begin{equation}
    c\left(\tilde{p}\right) = \min_{p \in P\left(u, v\right)} c(p)
\end{equation}
\subsection{First/Last mile Integrated Trip Planning} \label{sub: trip-plan}
\par Delhi's public transit network includes buses and metro. The bus network has 6700+ stops and 7000+ buses on 2000+ routes. The metro network consists of 250+ stops and 14 lines. Modeling the PT network using the approach described in section \ref{subsub:net-model} results in a dense network with 150,000+ edges. Traditional shortest path algorithms on such a graph are computationally expensive due the numerous exploration paths. To address this, we remodel the transit network as a bipartite graph, reducing the maximum path depth to 1 for any source-destination pair. This significantly reduces computational complexity, despite possibly having an equal or greater number of edges than the original graph $G$.
\subsubsection{Remodelled PTN Graph:} We take two sets $V^s$ and $V^d$ such that $V^s = V^d = V$.  The bipartite graph is then defined as $G_b(V^s \cup V^d, E_b)$ (Fig. \ref{fig:iPTN}(a)).  We define a directed edge $e_{ij} \in E_b$ for $i, j \in \{1, \ldots, m\}$ if: there exists a route between $v^s_i \in V^s$ and $v^d_j \in V^d$.  A tuple $(a_{ij}, b_{ij})$ is also associated with every edge $e_{ij} \in E_b$ where $a_{ij}$ and $b_{ij}$ are the travel time and the number of transfer between nodes $v^s_i \in V^s$ and $v^d_j \in V^d$ respectively. As the node $v^s_i = v^d_i$, edges $e_{ii}$ are not included in the graph.
\subsubsection{First/Last Mile Integration} To enable integrated trip planning that combines first/last mile and public transit services, we make two key observations. First, the route for first/last mile services depends on the origin and destination and cannot be pre-computed. Second, public transit connections between stops remain static and do not undergo regular changes.
\par
\begin{figure}[t]
    \centering
    \subfigure[]{\includegraphics[scale = 0.35]{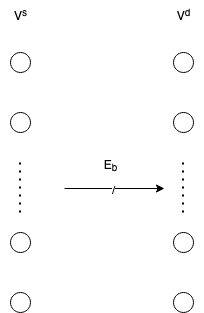}\label{fig:rmptn}}
    \hspace{0.05\columnwidth}
    \subfigure[]{\includegraphics[scale = 0.35]{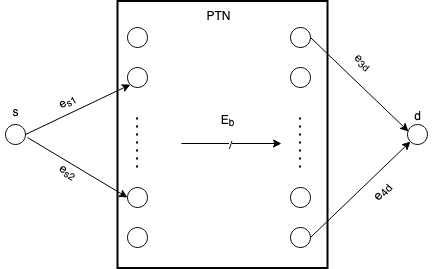}\label{fig:lmptn}}
    
    \caption{Public Transit Network Model: (a) Remodeled PTN Graph -- $V^s = V^d = V$. Directed edges $e_{ij} \in E_b$ are drawn from $V^s$ to $V^d$ iff a route exists between $v_i^s \in V^s$ and $v_j^d \in V^d$. (b) LMPTN Graph -- first/last mile edges $e_{si}, e_{di} \in E_{LM}$ are added to the graph in Fig.~\ref{fig:rmptn}.}
    \label{fig:iPTN}
\end{figure}

 Based on these observations, we propose a dynamic graph construction method that preserves the stationary component of the First/Last Mile Public Transit Network (LMPTN): the PTN graph $G_b$, for each source-destination pair. This eliminates the need to reconstruct the entire LMPTN for every new pair. The graph construction method follows these steps: Given a query, we introduce two dummy nodes, $s$ and $d$, into the graph $G_b$, representing the origin and destination locations, respectively. We define a new set of edges, $E_{LM}$, where directed edges $e_{si} \in E_{LM}$ connect $s$ to some $v_i^s \in V^s$, and $e_{id} \in E_{LM}$ connect some $v_i^d \in V^d$ to $d$. These edges represent the first/last mile trips to or from the transit mode stops. Consequently, we define the integrated first/last mile and public transit network graph (LMPTN) as $G_b(V^s \cup V^d \cup \{s, d\}, E_b \cup E_{LM})$ (Fig \ref{fig:iPTN}(b)).


\par To ensure practical first/last mile connections in a multi-modal transit scenario, we restrict the edges between the dummy source and destination nodes and nearby transit nodes by avoiding connections to distant locations. The distance range can be dynamically adjusted according to user preferences without compromising optimality.
    \subsubsection{Cost of a Path} In our path-cost definition (as in section \ref{subsub: path cost}), we account for multiple transit modes and acknowledge that users may have differing preferences among them. To reflect this, we introduce penalties for each transit mode. Let $w_{LM}$ and $w_{PT}$ represent the penalties for first/last mile travel time and public transit travel time, respectively. In a multi-modal network, a journey consists of two first/last mile legs and one public transit leg\footnote{To simplify definitions, we include the provision of walking whenever mentioning first/last mile, unless specified otherwise.}. For a path $p(s, d)$ between $(s, d)$, with $a \in V^s$ and $b \in V^d$ as the entry and exit points of the public transit leg, we define the following parameters:
\begin{enumerate}
    \item $tt_{PT}(a, b)$: the travel time through public transit between $a \in V^s$ and $b \in V^s$.
    \item $tt_{LM}(s, a)$ and $tt_{LM}(b, d)$: the travel time through a first/last mile service from the source to a transit stop $a \in V^s$ and from a transit stop $b \in V^s$ to the destination respectively. 
    \item  $f(p)$: the total fare of a multi-modal path $p$
\end{enumerate} 
 \par For a source destination pair $(s, d)$, we define a path $p\left(s, d\right)$  in $G_b$ as a vector $\left(s, a^*, b^*, d\right)$, where $a^* \in V^s$ and $b^* \in V^d$.  The cost of a path $c\left(p\right)$ is then a convex combination of the travel times incurred through first/last mile and public transit:
 \begin{equation}\label{eq: costlmpt}
  c\left(p\right) = c\left(s, a^*, b^*, d\right) = w_{LM}tt_{LM} +w_{PT}tt_{PT}(a^*, b^*) \end{equation}
where $w_{LM} + w_{PT} = 1$ and $w_{LM}, w_{PT} > 0$. 
\par Here, $tt_{LM}$ is the total time spent in a first/last mile service while travelling on path $p$ such that:
\begin{equation}
    tt_{LM} = tt_{LM}(s, a^*) + tt_{LM}(b^*, d)
\end{equation}
\subsubsection{Optimisation Problem}\label{subsub: opt-prob}
\par To obtain the optimal path between a source and destination, our objective is to find the transit stops $a^*$ and $b^*$ that minimise the time-cost $c\left(p\right)$ while satisfying the defined constraints. Let $MAX\_FARE$ be the upper limit of fare of the journey set by the user.
\par The optimisation problem is defined as:
\begin{align}
 \min_{a^* \in V^s \text{, } b^* \in V^d}& \quad c\left(s, a^*, b^*, d\right) \\
\text{s.t.} & \quad f(s,a^*, b^*, d) < MAX\_FARE \\ 
& \quad w_{LM} + w_{PT} = 1 \\ 
& \quad tt_{LM} = tt_{LM}(s, a^*) + tt_{LM}(b^*, d)\\
& \begin{aligned}
\quad w_{LM}, w_{PT}, f(s,a^*,b^*,d),&\\
\quad tt_{LM}(s, a^*), tt_{LM}(b^*, d)&> 0 
\end{aligned}
\end{align}

\par The solution to the above problem is a path starting from $s$ connecting to $a^*$ through a first/last mile, followed by a public transit connection from $a^*$ to $b^*$ and finally a first/last mile connection from $b^*$ to the destination $d$.
\section{Experiment Details}
\label{experiment}
\par We perform two experiments to analyse the effect of input parameters on path-efficiency and study the impact of Opti-Mile trips on city coverage.
\subsection{Experiment Parameters:}
\subsubsection{Input Parameters}\label{subsub:input-param}

Table \ref{tab:input-parameters} summarises the user-specific dynamic input parameters used in the experiments. The chosen values only serve to demonstrate the impact of input changes on the results of the optimisation problem in Section \ref{subsub: opt-prob}.


\begin{table}[t]
    \centering
    \caption{User-Specific Dynamic Input Parameters}
    \label{tab:input-parameters}
    \begin{tabular}{ll}
        \toprule
        Parameter & Values \\
        \midrule
        Max-Fare (Rs) & $50 + 10n$ for $n \in [0, 45]$ \\
        first/last mile Penalty ($w_{LM}$) & $0.1 + 0.1n$ for $n \in [0, 4]$ \\
        Public Transit Penalty ($w_{PT}$) & $1 - 2*w_{LM}$ \\
        LM Range ($r$ km) & $\{2, 5, 10\}$ \\
        \bottomrule
    \end{tabular}
\end{table}

\subsubsection{Path Attributes}\label{subsub:output-param}

Table \ref{tab:path-attributes} presents the attributes of the optimal path that were measured during the experiments.

\begin{table}[t]
    \centering
    \caption{Attributes of Optimal Path}
    \label{tab:path-attributes}
    \begin{tabular}{ll}
        \toprule
        Attribute & Description \\
        \midrule
        Fare/km & Average fare per kilometer of the trip \\
        Travel Time & Total time taken for the trip \\
        Total Distance & Distance traveled for the trip \\
        Public Transit Fare & Fare according to transit agency rules \\
        first/last mile Fare & Rs. 25 for first km, Rs. 10 for subsequent km \\
        \bottomrule
    \end{tabular}
\end{table}
\subsection{Experiment 1: Measuring the Efficiency of a Path}
\label{exp1: def}
\par For a path $p\left(u, v\right)$ between a source-destination pair $\left(u, v\right)$ in the LMPTN, efficiency $\Lambda\left(p\right)$ is a function of normalised convenience and cost-effectiveness of a path. Convenience, $C\left(p\right)$ is defined as the ratio of the total distance covered $d\left(p\right)$ to the cost of the path, while Cost-Effectiveness, $\left(E\left(p\right)\right)$ is the ratio of the total fare $\left(f\left(p\right)\right)$ to the total distance traveled:
\begin{alignat}{2}
C(p) &= \frac{d(p)}{c(p)}, \quad&\quad E(p) &= \frac{d(p)}{f(p)}
\end{alignat}

\par Since $E\left(p\right)$ and $C\left(p\right)$ have different units (kilometer/Rupee and km/s, respectively), we normalise them as $E_{norm}\left(p\right)$ and $C_{norm}\left(p\right)$ to facilitate a linear combination of the two factors\footnote{The maximum and minimum values of $E\left(p\right)$ and $C\left(p\right)$ are obtained from the results of our experiments for all the possible paths between a given source-destination pair.}.
\begin{equation}
    E_{norm}\left(p\right) = \frac{E\left(p\right) - \min\left(E\left(p^*\right)\right)}{\max\left(E\left(p^*\right)\right) - \min\left(E\left(p^*\right)\right)}
\end{equation}
\begin{equation}
    C_{norm}\left(p\right) = \frac{C\left(p\right) - \min\left(C\left(p^*\right)\right)}{\max\left(C\left(p^*\right)\right) - \min\left(C\left(p^*\right)\right)}
\end{equation}

\subsubsection{Efficiency of a Path}
\par The efficiency of a path $\Lambda\left(p\right)$ is the linear combination of $E_{norm}\left(p\right)$ and $C_{norm}\left(p\right)$:
\begin{equation}
    \Lambda\left(p\right) = w_C * C_{norm}\left(p\right) + w_E * E_{norm}\left(p\right)
\end{equation}
where $w_C, w_E > 0$ denote the weights assigned to convenience and cost-effectiveness, respectively, and $w_C + w_E = 1$.
\par A path $p_a$ is considered to be better than path $p_b$ if and only if $\Lambda\left(p_a\right) > \Lambda\left(p_b\right)$
\subsubsection{Methodology} \label{subsub: methodology}
\par The following steps were taken to perform the experiment:
\begin{enumerate}
    \item We first create the PTN graph mentioned in section \ref{sec: TP Overview}.
    \item 1000 pairs of source-destination locations were randomly sampled from Delhi-NCR region. For every source-destination pair:
    \begin{enumerate}
        \item Let the MAX\_FARE = 60, $w_{LM} = 0.2$, $w_{PT} = 0.8$, and LM Range = 5km.
        \item \label{pt1} Take graph $G_b$ and construct edges $e_{si}$ from $s$ to $v_i^s \in V^s$ where the distance between the stop $v_i^s$ and $s$,$ d^s\left(v_i^s, s\right) \leq \text{LM Range}$
        \item  Also construct edges $e_{jd}$ from $v_j^d \in V^d$ to $d$, where the distance between the stop $v_j^d$ and $d$, $d^d\left(v_j^d, d\right) \leq \text{LM Range}$
        \item In the resultant graph, solve the optimisation problem given in section \ref{subsub: opt-prob}
        \item \label{ptl} Record the attributes (Table \ref{tab:path-attributes}) of the optimal solution.
        \item Repeat from \ref{pt1} to \ref{ptl} for all combinations of the input parameter values given in Table \ref{tab:input-parameters}.
    \end{enumerate}
\end{enumerate}
\par The results from this experiment were used to make the observations detailed in section \ref{observations}.
\subsection{Experiment 2: Effect of Opti-Mile on Coverage}
\par
\begin{figure}[t]
    \centering
    \includegraphics[scale = 0.4]{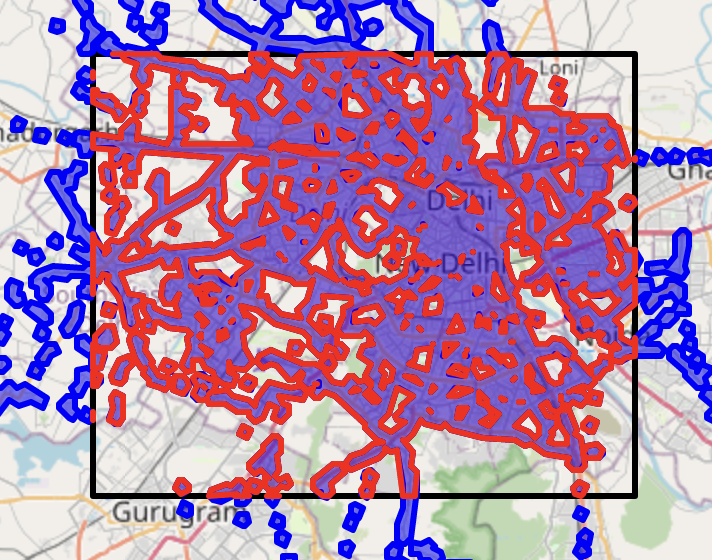}
    \caption{Coverage map showing the 90\% population density bounding box. Blue represents the transit network, while red represents the coverage area}
    \label{fig: coverage map}
\end{figure}

 To evaluate the coverage of the Public Transit Network (PTN) in Delhi, we employ a random sampling approach using a collection of bounding boxes on the map of the Delhi-NCR region. We establish four bounding boxes (Fig. \ref{fig: coverage map}) based on the boundaries that enclose the most densely populated municipal wards, representing the areas encompassing total population densities of 10\%, 50\%, 80\%, and 90\% of the total city population \cite{SEC}.
\par To analyse the impact of opti-mile trips on coverage, we calculate the combined area covered by the bus and metro network in Delhi. This coverage is then compared to the area covered through first/last mile trips. The coverage is measured for various walking or first/last mile distances, as indicated in Table \ref{tab: Coverage Table}.
\subsubsection{Transit network Coverage}
\par To measure the area coverage of the transit network, we perform the following steps for every bounding box:
\begin{enumerate}
    \item Mark the locations of all the bus and metro stops.
    \item For each stop, draw a circle with a radius of 500 meters. Exclude any area outside the bounding box to obtain the stop coverage area.
    \item Combine all the individual stop coverage areas to create a coverage map. 
    \item Calculate the total area coverage by dividing the area of the coverage map by the area of the bounding box.
\end{enumerate}
\subsubsection{Opti-Mile coverage} \label{subsub: optimile exp}
\par For every bounding box, we perform the following steps :
\begin{enumerate}
    \item \label{exp2:begin}Randomly sample 1000 random locations within the bounding box. For every location:
    \begin{enumerate}
        \item Identify the nearest bus and metro stops within a 2-kilometer range. These are the source stops.
        \item \label{exp2:pt1}For each source stop, determine all possible stops that can be reached through a direct connection and record their locations as destination stops.
        \item Create a circle with a radius of 2 kilometers around each source and destination stop. Exclude any area outside the bounding box to obtain the stop coverage area.
        \item Combine all the individual stop coverage areas to create a coverage map.
        \item \label{exp2:ptn}Calculate the area coverage by dividing the area of the coverage map by the area of the bounding box. This represents the coverage for one location.
        \item  Repeat steps \ref{exp2:pt1} to \ref{exp2:ptn} for different radius ranges and note the corresponding coverage values.
    \end{enumerate}
    \item The approximate opti-mile coverage for the bounding box is determined as the average of all the area coverage values recorded in \ref{exp2:begin}
\end{enumerate}
\section{Observations}
\label{observations}

\subsection{Experiment 1: Measuring the Efficiency of a Path}
\subsubsection{Fare and Distance Trade-Off}
\par An opti-mile trip involves trade-off between and distance. Table \ref{tab:farendist} hows that opti-mile trips have an 18\% higher median fare while covering 10\% less distance compared to optimal trips without transfer constraints, for the same source and destination. This trade-off may be attractive to non-regular users of public transit, who would otherwise rely on end-to-end cab services, as it offers a more efficient use of public transit without transfers.

\subsubsection{Factors Affecting Path Efficiency}
\par
\begin{figure}[b]
    \centering
    \includegraphics[width = 0.8\linewidth, scale = 0.8]{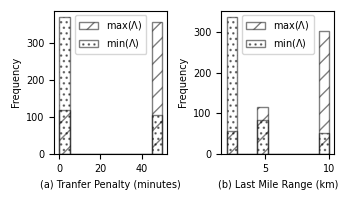}
    \caption{Factors Affecting Path Efficiency}
    \label{fig:integrated}
\end{figure}

 For some source-destination pairs, the paths with the highest and lowest $\Lambda$ may have the same fare even though the paths themselves may be different. In these cases, the $\Lambda$ values depends largely on the following factors:
\begin{itemize}
\item \textbf{Transfer Penalty:} Introducing a transfer penalty encourages the selection of opti-mile paths, i.e only direct public transit connections. We observed that 78\% of the paths with the highest $\Lambda$ preferred a high transfer penalty, while 71\% of the paths with low $\Lambda$ had no transfer penalty (Fig \ref{fig:integrated} (a)).
\item \textbf{First/Last mile Range:} From Fig. \ref{fig:integrated} (b), we observe that 83\% of the paths with the lowest $\Lambda$ have the lowest first/last mile range (2-3km), whereas the ones with higher $\Lambda$ prefer a larger first/last mile range ($>$ 3km) 90\% of the times. A larger range for first/last mile services implies longer distances traveled to reach a stop with a direct connection to the destination.
\end{itemize}

\par The findings described above support the notion that opti-mile journeys, with fewer transfers, sometimes even at the cost of longer first/last mile distances, are the most efficient paths.

\begin{table}[t]
\caption{Fare v/s Distance Comparison}
\label{tab:farendist}
\centering
\begin{adjustbox}{width=\columnwidth,center, scale =0.9}
\begin{tabular}{lllll}
\toprule
                    & \multicolumn{2}{c}{Median}          & \multicolumn{2}{c}{Deviation}     \\ \cmidrule(lr){2-3} \cmidrule(lr){4-5}
 & Opti-Mile & Non-Opti-Mile & Opti-Mile & Non-Opti-Mile \\ \midrule
Total Fare(Rs)      & 130.00 & 110.00 & 20.09 & 17.71 \\
Total Distance (km) & 21.71  & 24.29  & 8.05  & 7.72  \\ \bottomrule
\end{tabular}
\end{adjustbox}
\end{table}

\subsection{Experiment 2: Effect of Opti-Mile on Coverage}
\par From our analysis on the coverage provided by the transit network and opti-mile, we draw the following conclusions
\begin{enumerate}
    \item Opti-Mile trips, covering a maximum first/last mile range of 2km, exhibit a slight 4\% reduction in coverage for areas representing 10\% of the total population density. In contrast, for areas encompassing 90\% of the total population density, there is a noteworthy increase in coverage by over 8\%. This demonstrates that Opti-Mile trips do not lead to a significant coverage loss. Furthermore, expanding the first/last mile range to 5km results in more than 99.5\% coverage across all scenarios.
    \begin{table}[t] 
    \caption{PTN and Opti-Mile coverage comparison} 
    \label{tab:coverage} 
    \centering
    \begin{adjustbox}{width=\linewidth,center, scale = 0.8}
    \begin{tabular}{lllll}
    \toprule
    Population (\%Density) & PTN Coverage (\%) & \multicolumn{2}{l}{Opti-Mile Coverage (\%)} &  \\\cmidrule{1-4}
                           & 500m         & 2km                & 3km               &  \\ 
    \midrule
    10\%                   & 78.56        & 75.31              & 90.08             &  \\
    50\%                   & 68.11        & 68.49              & 84.41             &  \\
    80\%                   & 59.73        & 64.93              & 82.90             &  \\
    90\%                   & 55.67        & 60.71              & 80.12             & \\ 
    \bottomrule
\end{tabular}
    \end{adjustbox}
    \end{table}
    \begin{table}[t] 
    \caption{PTN coverage by population and walk distance}
    \label{tab: Coverage Table} 
    \centering
    \begin{adjustbox}{width=\linewidth,center, scale = 0.8}
    \begin{tabular}{lll}
    \toprule
    Population (\%Density) & Walk Distance (km) & PTN Coverage (\%) \\ \midrule
    10\% & 0.5 & 78.56 \\
    10\% & 1   & 97.40 \\
    50\% & 0.5 & 68.11 \\
    50\% & 1   & 91.67 \\
    80\% & 0.5 & 59.73 \\
    80\% & 1   & 83.24 \\
    90\% & 0.5 & 55.67 \\
    90\% & 1   & 80.78 \\ \bottomrule
    \end{tabular}
    \end{adjustbox}
    \end{table}
    \item Table \ref{tab: Coverage Table} shows a significant decrease in coverage ranging from 20\% to 35\% when the maximum walking distance is reduced to 500m. This reduction in coverage greatly affects the accessibility of the public transit network. Depending solely on public transit in such cases may leave certain areas with limited coverage. 
    \end{enumerate}
\section{Conclusion}
\label{conclusion}
\par Through our experiments, we demonstrate how opti-mile trips contribute to a cost-effective and efficient transportation network. first/last mile services for short-distance trips can reduce traffic congestion on longer routes, and further improvements in ride-sharing services can help alleviate this issue. Integrating first/last mile options also provides commuters with flexibility to choose transportation modes based on personal preferences.
\par Implementing opti-mile trips to enhance public transit accessibility in Delhi poses significant challenges. Extensive research on first/last mile availability \cite{li_vehicle_2021, shehadeh_fleet_2021, song_demand_2019}, can provide valuable insights and recommendations for policymakers. By leveraging these findings and combining them with our transit model, policymakers can make informed decisions to enhance the effectiveness and efficiency of public transit systems.  
\bibliographystyle{IEEEtran}
\bibliography{main}

\begin{thebibliography}{10}
\providecommand{\url}[1]{#1}
\csname url@samestyle\endcsname
\providecommand{\newblock}{\relax}
\providecommand{\bibinfo}[2]{#2}
\providecommand{\BIBentrySTDinterwordspacing}{\spaceskip=0pt\relax}
\providecommand{\BIBentryALTinterwordstretchfactor}{4}
\providecommand{\BIBentryALTinterwordspacing}{\spaceskip=\fontdimen2\font plus
\BIBentryALTinterwordstretchfactor\fontdimen3\font minus
  \fontdimen4\font\relax}
\providecommand{\BIBforeignlanguage}[2]{{%
\expandafter\ifx\csname l@#1\endcsname\relax
\typeout{** WARNING: IEEEtran.bst: No hyphenation pattern has been}%
\typeout{** loaded for the language `#1'. Using the pattern for}%
\typeout{** the default language instead.}%
\else
\language=\csname l@#1\endcsname
\fi
#2}}
\providecommand{\BIBdecl}{\relax}
\BIBdecl

\bibitem{bosona_urban_2020}
\BIBentryALTinterwordspacing
T.~Bosona, ``\BIBforeignlanguage{en}{Urban {Freight} {Last} {Mile}
  {Logistics}—{Challenges} and {Opportunities} to {Improve} {Sustainability}:
  {A} {Literature} {Review}},'' \emph{\BIBforeignlanguage{en}{Sustainability}},
  vol.~12, no.~21, p. 8769, Jan. 2020, number: 21 Publisher: Multidisciplinary
  Digital Publishing Institute. [Online]. Available:
  \url{https://www.mdpi.com/2071-1050/12/21/8769}
\BIBentrySTDinterwordspacing

\bibitem{silva_sustainable_2023}
\BIBentryALTinterwordspacing
V.~Silva, A.~Amaral, and T.~Fontes, ``\BIBforeignlanguage{en}{Sustainable
  {Urban} {Last}-{Mile} {Logistics}: {A} {Systematic} {Literature} {Review}},''
  \emph{\BIBforeignlanguage{en}{Sustainability}}, vol.~15, no.~3, p. 2285, Jan.
  2023, number: 3 Publisher: Multidisciplinary Digital Publishing Institute.
  [Online]. Available: \url{https://www.mdpi.com/2071-1050/15/3/2285}
\BIBentrySTDinterwordspacing

\bibitem{park_first-last-mile_2021}
\BIBentryALTinterwordspacing
K.~Park, A.~Farb, and S.~Chen, ``\BIBforeignlanguage{en}{First-/last-mile
  experience matters: {The} influence of the built environment on satisfaction
  and loyalty among public transit riders},''
  \emph{\BIBforeignlanguage{en}{Transport Policy}}, vol. 112, pp. 32--42, Oct.
  2021. [Online]. Available:
  \url{https://www.sciencedirect.com/science/article/pii/S0967070X21002341}
\BIBentrySTDinterwordspacing

\bibitem{ferguson_facilitating_2021}
\BIBentryALTinterwordspacing
B.~Ferguson and A.~Sanguinetti, ``\BIBforeignlanguage{en}{{FACILITATING}
  {MICROMOBILITY} {FOR} {FIRST} {AND} {LAST} {MILE} {CONNECTION} {WITH}
  {PUBLIC} {TRANSIT} {THROUGH} {ENVIRONMENTAL} {DESIGN}: {A} {CASE} {STUDY}
  {OF} {CALIFORNIA} {BAY} {AREA} {RAPID} {TRANSIT} {STATIONS}},''
  \emph{\BIBforeignlanguage{en}{Proceedings of the Design Society}}, vol.~1,
  pp. 1577--1586, Aug. 2021, publisher: Cambridge University Press. [Online].
  Available: \url{https://tinyurl.com/3uvh59bs}
\BIBentrySTDinterwordspacing

\bibitem{shaheen_mobility_2016}
\BIBentryALTinterwordspacing
S.~Shaheen and N.~Chan, ``\BIBforeignlanguage{en}{Mobility and the {Sharing}
  {Economy}: {Potential} to {Overcome} {First}- and {Last}-{Mile} {Public}
  {Transit} {Connections}},'' 2016. [Online]. Available:
  \url{https://escholarship.org/uc/item/8042k3d7}
\BIBentrySTDinterwordspacing

\bibitem{rahman_first-and-last-mile_2022}
\BIBentryALTinterwordspacing
M.~Rahman, M.~S. Akther, and W.~Recker, ``\BIBforeignlanguage{en}{The
  first-and-last-mile of public transportation: {A} study of access and egress
  travel characteristics of {Dhaka}’s suburban commuters},''
  \emph{\BIBforeignlanguage{en}{Journal of Public Transportation}}, vol.~24, p.
  100025, Jan. 2022. [Online]. Available:
  \url{https://www.sciencedirect.com/science/article/pii/S1077291X2200025X}
\BIBentrySTDinterwordspacing

\bibitem{goel_accessegress_2016}
\BIBentryALTinterwordspacing
R.~Goel and G.~Tiwari, ``\BIBforeignlanguage{en}{Access–egress and other
  travel characteristics of metro users in {Delhi} and its satellite cities},''
  \emph{\BIBforeignlanguage{en}{IATSS Research}}, vol.~39, no.~2, pp. 164--172,
  Mar. 2016. [Online]. Available:
  \url{https://www.sciencedirect.com/science/article/pii/S0386111215000278}
\BIBentrySTDinterwordspacing

\bibitem{grahn_optimizing_2022}
\BIBentryALTinterwordspacing
R.~Grahn, S.~Qian, and C.~Hendrickson, ``\BIBforeignlanguage{en}{Optimizing
  first- and last-mile public transit services leveraging transportation
  network companies ({TNC})},'' \emph{\BIBforeignlanguage{en}{Transportation}},
  Jun. 2022. [Online]. Available:
  \url{https://doi.org/10.1007/s11116-022-10301-z}
\BIBentrySTDinterwordspacing

\bibitem{xiong_optimal_2021}
\BIBentryALTinterwordspacing
J.~Xiong, B.~Chen, Z.~He, W.~Guan, and Y.~Chen,
  ``\BIBforeignlanguage{en}{Optimal design of community shuttles with an
  adaptive-operator-selection-based genetic algorithm},''
  \emph{\BIBforeignlanguage{en}{Transportation Research Part C: Emerging
  Technologies}}, vol. 126, p. 103109, May 2021. [Online]. Available:
  \url{https://www.sciencedirect.com/science/article/pii/S0968090X21001285}
\BIBentrySTDinterwordspacing

\bibitem{xie_solution_2010}
C.~Xie, H.~Gong, and F.~Wang, ``A solution for the last mile problem of the
  {Beijing} rapid transit network: {Local} shuttle bus system,'' in \emph{2010
  18th {International} {Conference} on {Geoinformatics}}, Jun. 2010, pp. 1--6,
  iSSN: 2161-0258.

\bibitem{pei_flexible_2019}
\BIBentryALTinterwordspacing
M.~Pei, P.~Lin, R.~Liu, and Y.~Ma, ``\BIBforeignlanguage{en}{Flexible transit
  routing model considering passengers’ willingness to pay},''
  \emph{\BIBforeignlanguage{en}{IET Intelligent Transport Systems}}, vol.~13,
  no.~5, pp. 841--850, 2019, \_eprint:
  https://onlinelibrary.wiley.com/doi/pdf/10.1049/iet-its.2018.5220. [Online].
  Available: \url{can you please provide the research objective, methodology
  and the conclusion of this web page #no_search}
\BIBentrySTDinterwordspacing

\bibitem{qiu_demi-flexible_2015}
\BIBentryALTinterwordspacing
F.~Qiu, J.~Shen, X.~Zhang, and C.~An, ``\BIBforeignlanguage{en}{Demi-flexible
  operating policies to promote the performance of public transit in low-demand
  areas},'' \emph{\BIBforeignlanguage{en}{Transportation Research Part A:
  Policy and Practice}}, vol.~80, pp. 215--230, Oct. 2015. [Online]. Available:
  \url{https://www.sciencedirect.com/science/article/pii/S0965856415002232}
\BIBentrySTDinterwordspacing

\bibitem{iliopoulou_metaheuristics_2019}
\BIBentryALTinterwordspacing
C.~Iliopoulou, K.~Kepaptsoglou, and E.~Vlahogianni,
  ``\BIBforeignlanguage{en}{Metaheuristics for the transit route network design
  problem: a review and comparative analysis},''
  \emph{\BIBforeignlanguage{en}{Public Transport}}, vol.~11, no.~3, pp.
  487--521, Oct. 2019. [Online]. Available:
  \url{http://link.springer.com/10.1007/s12469-019-00211-2}
\BIBentrySTDinterwordspacing

\bibitem{kumar_algorithm_2021}
\BIBentryALTinterwordspacing
P.~Kumar and A.~Khani, ``\BIBforeignlanguage{en}{An algorithm for integrating
  peer-to-peer ridesharing and schedule-based transit system for first
  mile/last mile access},'' \emph{\BIBforeignlanguage{en}{Transportation
  Research Part C: Emerging Technologies}}, vol. 122, p. 102891, Jan. 2021.
  [Online]. Available:
  \url{https://www.sciencedirect.com/science/article/pii/S0968090X20307919}
\BIBentrySTDinterwordspacing

\bibitem{jain_improved_2019}
S.~Jain and P.~Biyani, ``Improved {Real} {Time} {Ride} {Sharing} via {Graph}
  {Coloring},'' in \emph{2019 {IEEE} {Intelligent} {Transportation} {Systems}
  {Conference} ({ITSC})}, Oct. 2019, pp. 956--961.

\bibitem{SEC}
\BIBentryALTinterwordspacing
D.~State Election~Commission, ``{Delimitation 2022},'' accessed 2023-03-10.
  [Online]. Available:
  \url{https://sec.delhi.gov.in/sites/default/files/SCERT/circulars-orders/wardwise_population_summary.pdf}
\BIBentrySTDinterwordspacing

\bibitem{li_vehicle_2021}
\BIBentryALTinterwordspacing
C.~Li, D.~Parker, and Q.~Hao, ``\BIBforeignlanguage{en}{Vehicle {Dispatch} in
  {On}-{Demand} {Ride}-{Sharing} with {Stochastic} {Travel} {Times}},'' in
  \emph{\BIBforeignlanguage{en}{2021 {IEEE}/{RSJ} {International} {Conference}
  on {Intelligent} {Robots} and {Systems} ({IROS})}}.\hskip 1em plus 0.5em
  minus 0.4em\relax Prague, Czech Republic: IEEE, Sep. 2021, pp. 5966--5972.
  [Online]. Available: \url{https://ieeexplore.ieee.org/document/9636499/}
\BIBentrySTDinterwordspacing

\bibitem{shehadeh_fleet_2021}
\BIBentryALTinterwordspacing
K.~S. Shehadeh, H.~Wang, and P.~Zhang, ``\BIBforeignlanguage{en}{Fleet sizing
  and allocation for on-demand last-mile transportation systems},''
  \emph{\BIBforeignlanguage{en}{Transportation Research Part C: Emerging
  Technologies}}, vol. 132, p. 103387, Nov. 2021. [Online]. Available:
  \url{https://linkinghub.elsevier.com/retrieve/pii/S0968090X21003855}
\BIBentrySTDinterwordspacing

\bibitem{song_demand_2019}
\BIBentryALTinterwordspacing
Y.~Song, N.~Sun, and H.~Chen, ``\BIBforeignlanguage{en}{Demand {Adaptive}
  {Multi}-{Objective} {Electric} {Taxi} {Fleet} {Dispatching} with {Carbon}
  {Emission} {Analysis}},'' in \emph{\BIBforeignlanguage{en}{2019 {IEEE} {PES}
  {Asia}-{Pacific} {Power} and {Energy} {Engineering} {Conference}
  ({APPEEC})}}.\hskip 1em plus 0.5em minus 0.4em\relax Macao, Macao: IEEE, Dec.
  2019, pp. 1--5. [Online]. Available:
  \url{https://ieeexplore.ieee.org/document/8994675/}
\BIBentrySTDinterwordspacing

\end{thebibliography}
\addtolength{\textheight}{-12cm}   

\end{document}